\documentstyle[11pt,newpasp,twoside,epsfig]{article}
\makeatletter 
 
\def\gtrsim{\mathrel{\mathpalette\vereq>}} 
\def\vereq#1#2{\lower3pt\vbox{\baselineskip1.5pt \lineskip1.5pt 
\ialign{$\m@th#1\hfill##\hfil$\crcr#2\crcr\sim\crcr}}} 
\makeatother
\begin{document}

\title{THEORY AND EXTRAGALACTIC MASERS}
\author{William D. Watson} 
\affil{Department of Physics, 
University of Illinois, 
Urbana, IL 61801-3080}
\begin{abstract} 
Theoretical research on extragalactic masers (and hence this review) is 
almost entirely directed toward 22 GHz water masers in circumnuclear disks 
with the focus being the near ideal masing disk at the nucleus of the 
galaxy NGC4258. The discussion here is organized around (1) the excitation 
and conditions for the masers, (2) the spatial and spectral appearance of 
the maser emission, and (3) the mass accretion rate and structure of the 
disk. In addition, a summary is given of the basic physics (spectral 
linebeadths and maser polarization) that underlies certain interpretations 
of these and other astrophysical masers.%
\footnote{\em to appear in the proceedings of IAU Symposium \#206: Cosmic Masers\\
(Mangaratiba, Brazil; March 2001)} 
\end{abstract}
\vspace{-.25in}
\section{INTRODUCTION}

Although there are some thirty galaxies in which H$_2$O masers are observed 
(certain others have OH masers), there is little in the way of ``theory" other 
than for the few where there is evidence for disk structure. Even for 
these, almost all of the theoretical activity has been directed toward 
NGC4258 where the highly refined observations of 22 GHz water masers in a 
near ideal, thin Keplerian disk provide the best opportunity for applying 
and testing ideas. My discussion of the ``Theory and  Extraglactic Masers" 
is thus largely a summary of theory directed toward the masers in a single 
galaxy---but it is a paradigm of considerable importance, and the ideas 
involve the basic description of molecular accretion disks near active 
galactic nuclei. Analysis of the structure of this disk utilizing 
information from maser observations potentially provides opportunities to 
assess the mass accretion rate which, when related to the energy emitted 
from this nucleus in other wavebands (X-ray, infrared), bears on current 
issues concerning active galactic nuclei---especially, on the possibility for 
advection dominated accretion flows. Such analysis also may help clarify 
the nature of the long sought mechanism for viscosity in accretion disks. 
Further, the masing disk in NGC4258 provides arguably the best evidence for 
massive black hole at the nucleus of an active (or at 
least mildly active) galaxy. See Moran, Greenhill, \&  Herrnstein (1999)
for a summary of
the observational 
data about extragalactic H$_2$O masers.

\vspace{-10pt}
\section{THE ENVIRONMENT AND EXCITATION OF CIRCUMNUCLEAR WATER MASERS}

Masing by the water molecule at 22 GHz involves states (6$_{16}$ and 
5$_{23}$) which lie some 600K above the ground state. Elevated temperatures 
and relatively high gas densities are thus required to create strong 22 GHz 
masers. Gas kinetic temperatures of 400K and hydrogen densities of 10$^9$ 
cm$^{-3}$ are considered representative for galactic water masers, though 
somewhat lower temperatures (300K) and densities (10$^8$ cm$^{-3}$) may be 
adequate in circumnuclear disks where the dimensions of the masing regions 
can be larger. Since the earliest calculations (Schmeld, Strelnitskii, \& 
Muzylev 1976), galactic water masers other than the circumstellar variety 
are generally believed to be associated with shock waves. Shock waves 
almost certainly create some of the environments for the extragalactic 
masing of water, as well. However, the accretion disks are likely to be 
relatively quiescent. Further, strong X-ray emission is typical at the 
nuclei of these galaxies. Hence, the elevated temperatures in the gas of 
the accretion disk that are necessary for masing have seemed most likely to
be a result of heating by X-rays, though shock waves (Maoz \& McKee 1998)
and even 
viscous heating (Desch, Wallin, \& Watson 1998) cannot readily be  
dismissed.

The chemistry and excitation for masing in a dense gas on which 
X-rays with intensities that are likely in circumnuclear environments has 
been computed by Neufeld, Maloney \& Conger (1994). As in galactic water 
masers, thermal collisions elevate the water molecules to a range of 
excited states. The inversion results from the ensuing 
radiative/collisional cascade. Escape of a fraction of the infrared, 
spectral line radiation that is involved in the cascade is essential to 
prevent the thermalization of the populations of the masing states, and 
hence the quenching of the masing. In standard calculations including those 
of Neufeld et al. (1994), the escape (and hence the maser power) is limited 
by the trapping due to reabsorption of the infrared radiation by the water 
molecules themselves (see Figure 1). However, the maser power can be 
increased significantly by utilizing Deguchi's (1981) recognition that 
absorption of the infrared radiation by dust grains can serve as an 
alternative to escape for preventing the thermalization of the masing 
transition. The results of calculations (Collison \& Watson 1995) for the 
pumping that include the influence of this absorption by dust grains in the 
circumnuclear masing environments created by X-rays are shown in Figure 1. 
It is essential that the temperature of the grains remain below that of the 
gas. This requirement is satisfied for a wide range of likely conditions 
(Desch et al. 1998). Although the potentially important issue of 
time variations in the flux from extragalactic masers is beyond the scope 
of this discussion, it is noteworthy that changes in the temperatures of 
the dust grains may play a role in understanding the short time scales for 
the changes (Neufeld 2000; Gallimore et al. 2001).
\begin{figure} 
\begin{center} 
\epsfig{file=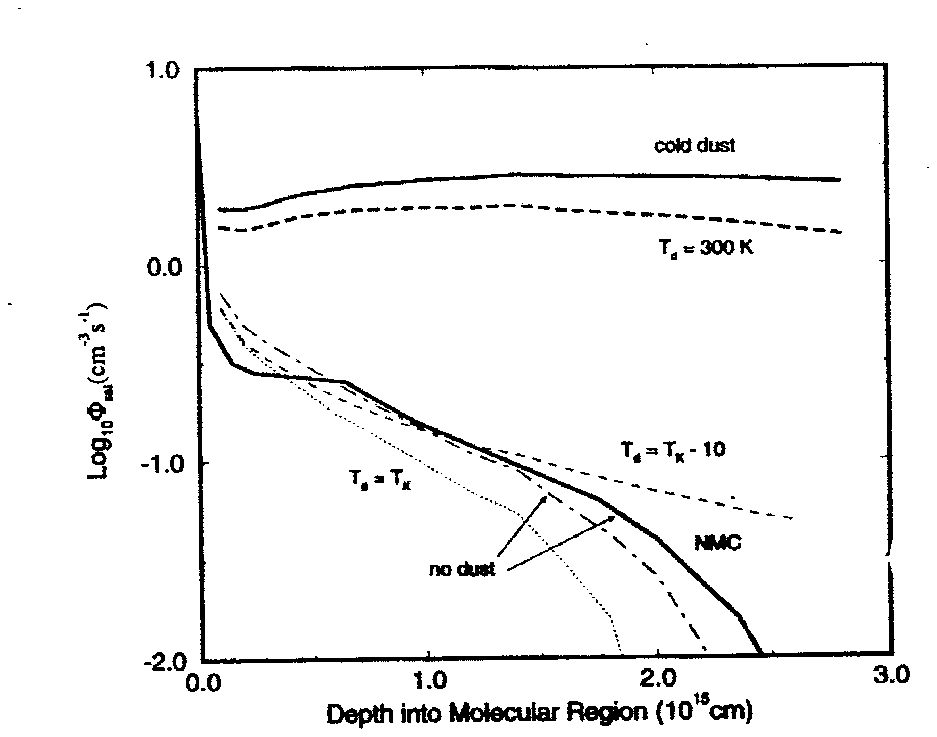, width=5in}
\caption{\footnotesize The log of the maser photon emissivity vs distance 
into a slab of gas that is irradiated by an X-ray flux similar to that 
expected for the circumnuclear masing gas in the galaxy NGC4258. The 
expected enhancement due to dust grains is best seen by comparing the lines 
labelled ``cold dust" and ``T$_d$ = 300K" with the solid line labeled NMC 
(Neufeld, Maloney, \& Conger 1994) from the calculation in which the 
effects of dust are not included. From Collison \& Watson (1995; also 
Watson \& Wallin 1997).} 
\end{center}
\end{figure}
\nopagebreak[4]
\vspace{-10pt}
\section{MASING IN A KERPLERIAN DISK AND ITS RELATIONSHIP TO
THE OBSERVATIONS}

>From the viewpoint of a distant observer in the plane of a masing Keplerian 
disk that is uniformly pumped in a ring between inner radius R$_i$ and 
outer radius R$_o$, the optical depths for unsaturated masing are largest 
toward the center of the disk and for rays at the sides of the disk with 
impact parameters that are near (but slightly less than) the outer radius 
of the disk. The gradients in velocity are a minimum in these directions. 
The emission at the sides arises mainly from locations along the rays where 
the rays are nearly tangent to the Keplerian orbits (i.e., near the midlines). 
Thus, three symmetrical peaks in the spectrum and three images extended 
along a line in the sky are considered indicative of a disk. If the masing 
is highly saturated, the optical depths at the sides will be largest for 
rays with impact parameters near the inner radius of the masing ring and 
will occur at Doppler velocities that are essentially equal to the 
velocities of the Keplerian orbits that are tangential to these rays.
Toward the center, the Doppler velocities of the most intense rays at each 
impact parameter will vary linearly with the impact parameter. This 
variation is proportional to the rotational velocity at the outer edge of 
the ring if the masing is unsaturated, and is proportional to the rotational 
velocity at the inner edge if the masing is highly saturated. In addition
to exhibiting
characteristics similar to the foregoing, the Doppler 
velocities of a number of microfeatures within the central emission of 
NGC4258 are observed to change linearly with time (Baan, Haschick, \&
Peng 1993).
As a result, the diverse data from NGC4258 can be interpreted as defining a
masing disk 
viewed nearly edge-on and the mass of the object at its center can be
inferred (Watson \& 
Wallin 1994). In more recent 
observations, images of these masing features have been observed to move 
across the sky. For NGC4258, the maser amplifications toward the center and 
at the sides are not at all similar. Emission toward the center is 
prominent, but only because the masing here is amplifying a strong 
continuum source at the center. Clearly, the clumpy appearance of the maser 
emission from NGC4258 indicates some form of irregularities in the medium 
of the disk. At the same time, the medium must be smooth enough (or the 
irregularities numerous enough) that emission at the sides occurs mainly 
near the midlines where the lengths for coherence in velocity are greatest. 
Widely separated, but ``aligned" masers can be quite intense (Deguchi \& 
Watson 1989).

Turbulence is prevalent in astrophysical environments of diffuse 
matter (e.g., supersonic turbulence in molecular clouds). MHD turbulence 
is believed to be involved in the viscous dissipation mechanism for 
accretion disks. It is thus natural to inquire as to whether the rather 
gentle irregularities in the optical depths that are caused by turbulence 
of a ``generic" form are sufficient to create spectra that are similar to 
those that are observed for NGC4258. We have thus performed computations 
 that utilize standard methods to create representative turbulent velocity 
fields in the gas of a disk with dimensions and with Keplerian velocities 
similar to those of the NGC4258 disk. For simplicity, the maser opacity is 
taken as constant and maser saturation is ignored. Examples of spectra from 
our computations are given in Figures 2 and 3 for rms turbulent velocities 
that are equal to the sound speed. What is shown in these Figures is the 
maser amplification due to the disk. We have reasoned that the magnitudes 
of the amplification factors in these Figures, when combined with best 
estimates for the intensities of the background continuum radiation that is 
being amplified, are approximately what is required to yield the observed 
fluxes. These spectra can be compared with observations in, e.g., Myoshi et 
al. (1995).Rms turbulent velocities that are approximetely equal to the 
sound speed in the masing gas
seem to be necessary to obtain spectra that are similar to the observed
spectra. 
Deguchi (1982) originally demonstrated that representative 
turbulent velocity fields obtained in a somewhat analogous way lead to 
images for circumstellar masers that are similar to the observed images 
(also, Sobolev, Wallin, \& Watson 1998 for spectra and images of methanol
masers in a turbulent medium).
\begin{figure} 
\epsfig{file=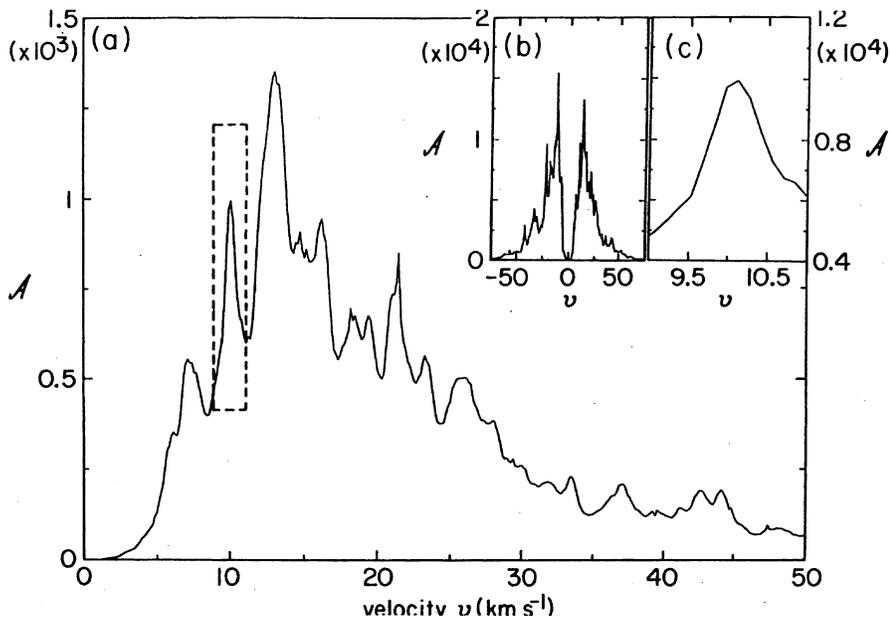, width=5in}
\begin{center} 
\caption{\footnotesize A representative spectrum from the simulations of a 
turbulent, masing Keplerian disk that is intended to describe the central 
emission from the circumnuclear disk in NGC4258. (a) An expanded view to 
show the microstructure of the full spectrum in panel (b). The feature 
within the dashed rectangle is expanded further in panel (c) to exhibit the 
narrow breadth of the feature. The spectrum (versus Doppler velocity) is
presented in terms of the factor by which the background continuum source
is amplified by the masing. From Wallin, Watson, \& Wyld (1998).} 
\vspace{-.3in}
\end{center}
\end{figure}

\nopagebreak[4]
\begin{figure}
\epsfig{file=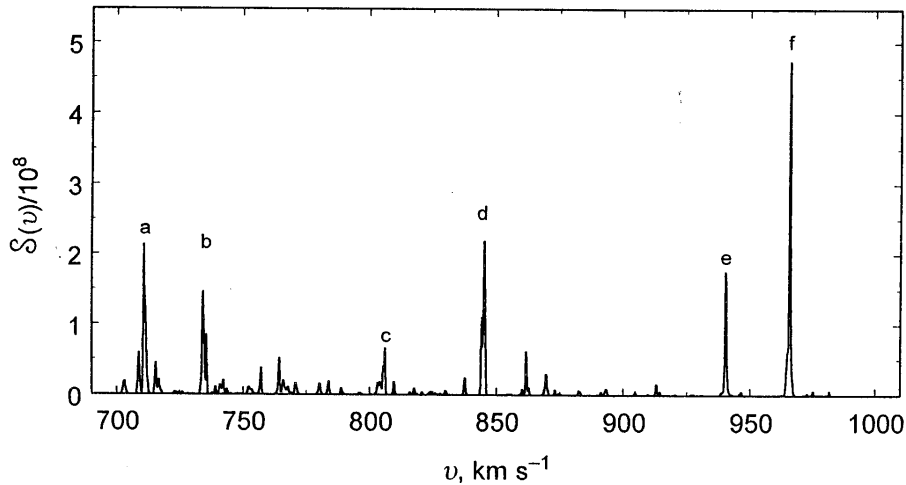, width=5in}
\begin{center}
\caption{\footnotesize Similar to Figure 2 except that the spectrum here is 
intended to describe the maser emssion from the side of the disk. The 
meaning of the amplification factor is analogous to that in Figure 2 though 
its magnitude is quite different because these masers are not amplifying a 
strong continuum source. From Wallin,Watson \& Wyld (1999).} 
\end{center} 
\end{figure}
The possibility for a quite different origin of the masing features 
at the sides of the disk has been recognized by Maoz \& McKee (1998). The 
reasoning in this interpretation is that the mass of the disk may be within 
the range where the Toomre stability parameter $Q$ indicates that spiral 
density waves will form. The features at the sides are then proposed to 
occur where the lines of sight are tangent to these spirals. 
Shocks form at the edges of these 
waves where gas flows into the wave, and provide the conditions for the 
water masers. Because the pattern of the waves is rotating, the locations 
where the tangents occur drift outward in radius to lower Keplerian 
rotational velocities. An acceleration of the Doppler velocities of the 
masing features is thus predicted. The absence of such accelerations in 
refined observations (Bragg et al. 1999) is in conflict with the proposed 
interpretation---at least for what seem to be the most likely choices for 
the relevant parameters. In a somewhat related idea but directed toward the 
presumed masing disk of NGC1068, Kumar (1999) describes how gravitational 
instabilities in the disk can create the masing clumps when the mass of the 
disk is larger than that for which spiral density waves are expected. Other 
discussions of the nature of the masing disk include Kartje, Konigl, \&
Elitzur (1999), and Babkovskaya \& Varshalovich (2000). Novel aspects of
radiative processes also have been examined (Deguchi 1994; Gangadahara,
Deguchi, \& Lesch 1999).

A pervasive issue in understanding and interpreting the observations 
of astrophysical masers is the degree of saturation. To proceed directly 
from the observed flux, one must resort to highly uncertain estimates for 
the beaming angle---ordinarily utilizing a cylindrical geometry for the 
masing region. For Keplerian disks, the cylindrical approximation tends to 
overestimate the degree of saturation (Watson \& Wyld 2000). Nevertheless, 
we and others have reasoned in this way that the masers in NGC4258 are 
likely to be no more than partially saturated (see Wallin et al.1998). An 
alternative approach is to recognize  
that (in the absence of velocity gradients) the spectral linebreadths of 
unsaturated masers are narrower than the thermal breadths, and that the 
spectral lines rebroaden to the full thermal breadths when the maser becomes 
saturated. Despite assertions to the contrary, our own rigorous 
calculations (Emmering \& Watson 1994) demonstrate that the rebroadening 
does occur to the full Maxwellian breadth for spherical as well as for 
linear masers---and hence, almost surely this result is independent of 
geometry. At the gas kinetic temperature of 400K that is ordinarily adopted 
as the approximate minimum for the pumping of 22 GHz masers, the thermal 
linebreadth (FWHM) is 1 km s$^{-1}$ for water molecules. I believe that it 
is far more likely that the 22 GHz line is actually a merger of the three 
strongest hyperfine components of the 6$_{16}$-5$_{23}$ transition. They 
merge so well that the line profile can be undistorted. With the three 
merged hyperfine components, the minimum breadth for the fully rebroadened 
22 GHz maser line is 1.5 km s$^{-1}$ (Nedoluha \& Watson 1991). Not only 
are the linebreadths of some of the galactic water masers with the highest 
fluxes less than 1.5 km s$^{-1}$, they are even less than 1 km s$^{-1}$! 
The prominent 1306 km s$^{-1}$ feature from the side of the disk of 
NGC4258 has a linewidth of 1.1 km s$^{-1}$. Many other microfeatures have 
breadths less than 1.4 km s$^{-1}$ (Bragg, et al. 2000), and some are even
less than 1 km s$^{-1}$.
Note that velocity gradients, 
unresolved components, etc. can increase the linebreadth, but no valid way is 
known to reduce the breadth other than unsaturated amplification. 
\begin{figure}[t]
\epsfig{file=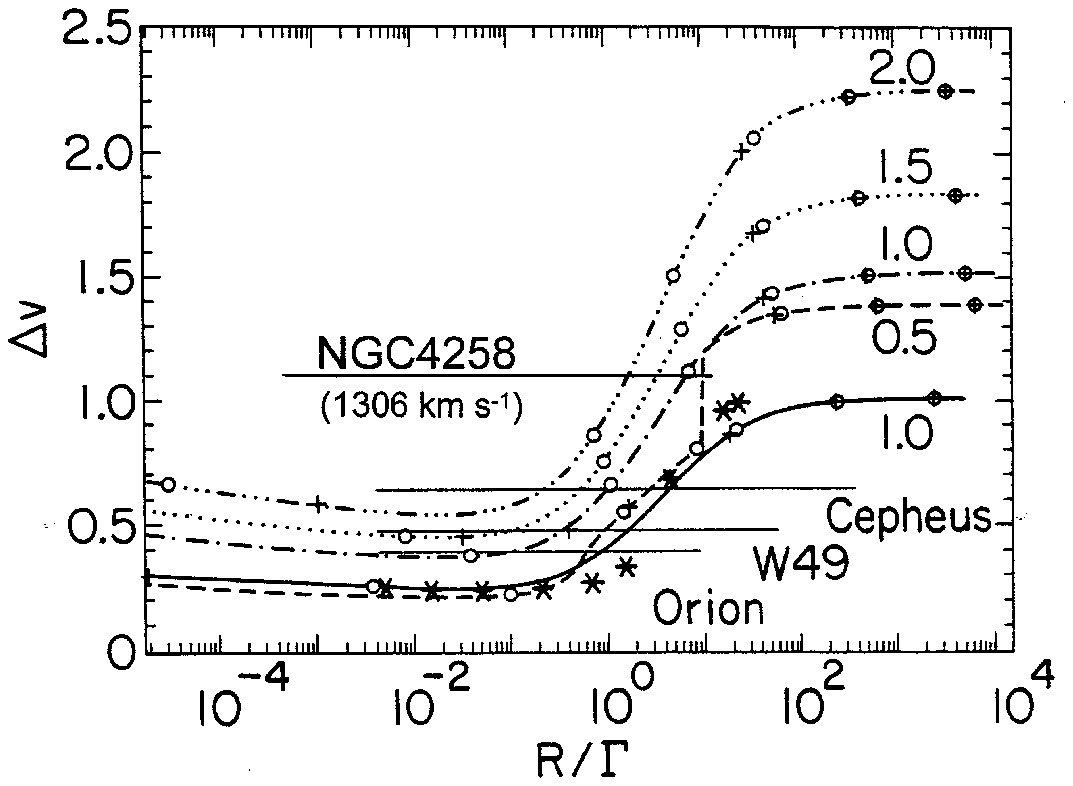, width=5in} 
\begin{center} 

\caption{\footnotesize  The linebreadth (FWHM) versus the degree of saturation 
for maser spectral lines. The solid line is computed for a transition 
between two molecular states, whereas the other lines are computed for the 
6$_{16}$-5$_{23}$ (22 GHz) transition of water that is assumed to be the 
result of the merger of the three strongest hyperfine components. The lines 
are labeled according to the thermal breadth of the Maxwellian distribution 
of the molecular velocities (note that 1 km s$^{-1}$ corresponds to 400K). 
The computations represented by lines are performed in the idealization 
that the masers are linear (from Nedoluha \& Watson 1991). Also shown and 
indicated by asterisks are the results of similar computations for a 
two-level, spherical maser which demonstrate that the spectral line 
rebroadens in the same manner for this geometry (from Emmering \& Watson 
1994). The horizontal lines indicate the linebreadths for some of the most 
intense flares of galactic, 22 GHz water masers and for the prominent 1306 km 
s$^{-1}$ feature from the side of the masing disk in NGC4258.} 
\vspace{-.5in}
\end{center}
\end{figure}

The relaxation of molecular velocities, mainly through the emission and 
absorption of trapped, infrared spectral line radiation (Goldreich \& Kwan
1974) is a way 
to understand how masers can become saturated while retaining the narrow 
linebreadth of unsaturated masers. Then, the rebroadening of the spectral 
line is postponed until the rate for stimulated emission exceeds the rate 
for velocity relaxation. When velocity relaxation is important for H$_2$O 
masers, the decay rate $\Gamma$ is reduced due to the trapping. However,
the rate 
for velocity relaxation is approximately the  value that ordinarily is 
adopted for $\Gamma$---the Einstein A-value for de-excitation of the 
molecular states by infrared emission (approximately 2 s$^{-1}$ for the 22 
GHz states). Thus, the numerical value of the rate for stimulated emission 
at which rebroadening occurs is essentially unchanged as a result of 
velocity relaxation. Though this has been established in detail only for 
the 22 GHz masing transition (Anderson \& Watson 1993), I expect that it is 
a rather general result.
\vspace{-10pt}
\section{THE ACCRETION RATE AND STRUCTURE OF THE MASING DISK}

A fraction (ordinarily taken as about five percent) of the rest energy 
of the matter that falls onto a black hole emerges as the bolometric 
luminosity in a standard interpretation. Typically, some ten percent of the
bolometric luminosity appears as X-rays. By assuming steady-state, the
accretion 
rate through the disk in NGC4258 can be estimated as $\dot{M} \approx 1.4 
\times 10^{-4} M_\odot yr^{-1}$. An outstanding issue for the 
``underluminous" nuclei in mildly active galaxies such as NGC4258 is 
whether this estimate is valid and the accretion rate is lower than for the 
more luminous, active galactic nuclei. Alternatively, the accretion rates 
may be similar to those in the luminous nuclei, but much of the energy 
fails to appear in the bolometric luminosity because it is advected into 
the black hole (advection dominated accretion flows, or ADAF's). The refined 
description of the masing disk in NGC4258 offers a possibility that 
valuable information can be obtained on this issue. Data at various 
wavelengths has been fit to an assumed ADAF spectrum for NGC4258 and leads 
to $\dot{M} \approx 10^{-2} -10^{-3} M_\odot yr^{-1}$ (Lasota et al.1996; 
Gammie, Narayan, \& Blandford 1999; Chary et al. 2000). From the 
observations of the masing disk, Maoz \& McKee (1998) obtain $7\times 
10^{-3} M_\odot yr^{-1}$ in the spiral density wave interpretation 
whereas Neufeld \& Maloney (1995) find $7\times 10^{-5} \alpha M_\odot 
yr^{-1}$ based mainly on the premise that the masing extends outward in radius 
until the molecules disappear and the gas becomes atomic. Here, $\alpha$ is
the 
coefficient from the Shakura-Sunyaev description of viscosity. An approach 
that seems somewhat more direct involves determining the strength of the 
magnetic field in the disk (or a useful upper limit) by observing for the 
polarization of the maser radiation. Assuming a balance between the thermal 
and magnetic pressure within the disk, together with the Shakura-Sunyaev 
description of turbulence, then leads to a value for 
($\dot{M}/\alpha$). Herrnstein et al. (1998) thus infer an upper limit 
$10^{-2} \alpha M_\odot yr^{-1}$ from an upper limit of 300 mG for the 
magnetic field based on their failure to detect circular polarization at a 
level of 1.5 percent in the 1306 km s$^{-1}$ feature at the side of the 
disk. Though this limit is not sensitive enough to discriminate between an 
ADAF and a standard model, the prospects are good for observations of the 
central spectrum that are considerably more sensitive (see also 
Moran 2000). However, the spectrum for the circular polarization of the 
central emission will be quite different from that of a two-level 
transition if the 22 GHz line is a merger of three hyperfine components 
(Watson \& Wiebe 2001) as I believe is most likely. Relating the magnetic 
field strength to the circular polarization probably is less difficult for 
isolated spectral lines (see also Nedoluha \& Watson 1992).

The mass in the disk can be related to $M_\odot$ utilizing the 
Shakura-Sunyaev prescription for the viscosity. In terms of the radius 
$R_i$ at the location of the central masers, the surface mass density at
radius 
R is (e.g., Desch et al. 1998)
\begin{equation} 
\Sigma\ =\ (20/\alpha )(\dot{M}/2\times10^{-4}M_\odot 
yr^{-1})(400K/T)(R_i/R)^{3/2}\ {\rm g\ cm^{-2}} 
\end{equation} 
Since the X-rays from the central source can penetrate to only about ${\rm
1\ g\ 
cm^{-2}}$, they probably fail by a considerable factor in being able to 
maintain the 400K or so temperatures that would be required 
for masing by water throughout the disk---even for the $\dot{M}$ based on
the ``standard" 
interpretation of flow through the accretion disk. Thus, a cold component 
of the disk probably exists that is more massive than 
the masing component that we detect. The spectrum of NGC4258 often exhibits 
a narrow dip near the systemic velocity. This may be due to absorption by
the cold component. 
Beyond the depths to which the 
X-rays can penetrate and excluding isolated regions where there might be 
localized heating, the temperature probably is less than about 50K. With 
favorable assumptions about poorly understood processes, viscous heating 
within the disk might be sufficient to heat a thin layer that could be the 
site of the observed masing (Desch et al. 1998). It could not, however, 
provide sufficient energy to heat the entire thickness of the disk to such 
temperatures. Note that heating by shocks in spiral density waves is 
included in this assessment of the viscous heating since the origin of the 
energy for both is the change in gravitational potential energy as the 
matter drifts inward. An estimate for the mass in the disk between the 
inner and outer radii at which masers are observed can readily be obtained 
by ignoring the variation in temperature in the above expression for 
$\Sigma$ and integrating to obtain 
\begin{equation} 
M_{disk} \simeq (10^4 /\alpha )(\dot{M}/2\times 10^4 M_\odot 
yr^{-1})(400K/T)M_\odot 
\end{equation}

The central mass about which the masing disk is orbiting has a mass of 
$3.6\times 10^7 M_\odot$. I am not aware of a quantitative assessment for 
the maximum disk mass that is consistent with the absence of observed 
deviations from the Keplerian rotation, but it seems unlikely to be larger
than 
about 0.3 percent of the central mass. Based on the above M$_{disk}$, there 
is then little if any flexibility for $\dot{M}$ to exceed the minimum 
value inferred from the standard description of accretion. Gravitational 
clumping and/or multiple thermodynamic phases may alter the simple analysis 
given above in essential ways. Certainly, the Shakura-Sunyaev prescription 
for viscosity is an extreme simplification. For example, MHD simulations 
tend to indicate that the magnetic pressure (and hence the likely 
dissipation of viscous heat) decreases much less rapidly with distance away 
from the midplane of an accretion disk than does the gas kinetic pressure 
(Miller \& Stone 2000), a viewpoint that has received support from other 
perspectives (e.g., Stella \& Rosner 1984). Finally, by utilizing the 
Shakura-Sunyaev viscosity once more, the Toomre stability parameter $Q$ can 
be expressed as
\begin{equation} 
Q\simeq 7\alpha (T/400K)^{3/2} (2\times 10^{-4} M_\odot yr^{-1}/\dot{M})
\end{equation}

A disk is stable for $Q\geq1$. This marginal stability (at best) of the 
disk in NGC4258 underlies proposals for density waves and for the clumping
of the
mass. We
note that an entirely different, alternative approach exists for
understanding accretion
in disks (Blandford \& Payne 1982), but does not lend itself to quantitative
interpretations.
\vskip.2in 
 
\noindent{\bf APPENDIX---POLARIZATION OF MASER RADIATION}
\vskip.1in
The contributions to issues of broad astrophysical importance from the 
observation of extragalactic masers, just as for galactic masers, often 
depend upon an understanding of the basic physics for the excitation and 
transport of astrophysical maser radiation. This is especially true in 
efforts to infer information about the magnetic field from the observation 
of the polarization of the maser radiation. Confusion about the theory for 
maser polarization seems to be prevalent in the literature. A summary at 
this time may thus be helpful. I will focus here only on the limit in which 
the Zeeman splitting $g\Omega$ is much smaller than the spectral linebreadth 
$\Delta\omega$ (``weak splitting"). This is the limit that is relevant for the 
circular polarization of extragalactic water masers, as well as for 
galactic H$_2$O, SiO, OH 1720 MHz, and CH$_3$OH masers. For simplicity, I will 
ignore any explicit consequences of cross-relaxation in this discussion.
Cross-relaxation is analogous to velocity relaxation in that its main effect
for polarization in the weak splitting regime is that the rate for
cross-relaxation
$\gamma$ replaces
the "phenomenological decay rate" $\Gamma$ of the molecular populations in the 
discussion of this Appendix (Goldreich, Keeley, \& Kwan 1973b). However, as
in the case
of velocity relaxation, $\Gamma$ ordinarily is reduced when cross-relaxation 
is important
and $\gamma$ has the approximate value that normally is associated with
$\Gamma$---approximately 2 s$^{-1}$ for the 22 GHz water masers.

\vskip.2in 
\noindent {\bf A. The Idealization of Goldreich, Keeley and Kwan: Linear
Polarization}
\vskip.1in

A discussion of the theory of maser polarization naturally begins with the 
classic paper of Goldreich, Keeley, \& Kwan (1973a; hereafter GKK). It is
therefore 
important to know exactly what idealized masing conditions were considered 
by GKK, and what answers they did and did not provide. GKK considered the 
basic issue of the polarization of maser radiation that emerges from a 
linear maser when weak, continuum seed radiation is incident at the far end 
of a masing region in which there is a constant magnetic field. They 
considered only an angular momentum J=1-0 masing transition and only equal 
pumping of the magnetic substates of J=1. I will refer to this scenario as 
the ``GKK idealization". {\it The key restriction in GKK is that the 
solutions for the polarization given there are obtained only in limiting 
regimes for the maser intensity, as specified by the rate R for stimulated 
emission.} With $\Gamma$ as the usual decay rate for 
the masing states, GKK find that there is linear polarization when 
(a)$\Gamma \ll R \ll g\Omega$ or when (b) $g\Omega \ll R \ll (g\Omega)^2 
/\Gamma$, and that the fractional linear polarization is zero when 
$R\ll\Gamma$ or $(g\Omega )^2/\Gamma \ll R$. In regime (a), the fractional 
linear polarization (Q/I) is given by the widely quoted expressions (Q/I) = 
-1 for angles $\theta$ between the magnetic field and the line-of-sight 
that are less than 35 degrees and for other angles between 
35 and 90 degrees, ${\rm (Q/I)= 3 sin^2 \theta - 2)/3 sin^2\theta}$. In
regime (b), Q/I is positive and can be as large as one-third. Positive Q
corresponds to a net linear polarization that is perpendicular to the
magnetic field.

Although it accomplished a lot, the GKK analysis was incomplete even for 
the GKK idealization because it did not tell us quantitatively how strong 
the inequalities that define the various regimes must be in order for the 
solutions given there for the linear polarization to be applicable. More 
generally, the behavior of the polarization as a continuous function of 
maser intensity (or degree of saturation) was not provided. Several 
collaborative efforts at Illinois have largely completed the picture for 
the GKK idealization. I will now summarize the key results.

Computations for the linear polarization as a continuous function of maser 
intensity were initially limited to intensities for which $R \ll g\Omega$ 
(Western \& Watson 1984). As expected, the computations agreed with the 
GKK solutions in the limit of high saturation $(\Gamma \ll R)$. However, 
quite high saturations $R/\Gamma \approx 100$ are necessary to attain the 
fractional linear polarizations of 70 percent for J=1-0 transitions which 
are sometimes observed for the SiO masers. To attain the comparably high 
fractional polarization that is observed for masing of SiO in the J=2-1 
transition, $R/\Gamma$ must be implausibly large. It follows
that 
linear polarization cannot be entirely due to the GKK mechanism. The next 
step was rephrasing the GKK density matrix equations in a way that is 
convenient for treating more generally the regime where $R\ll g\Omega$ is 
not satisfied. An additional result in this investigation was the 
recognition that significant linear polarization is created in regime (b) 
[where $g\Omega \ll R \ll (g\Omega)^2/\Gamma$] only for a J=1-0 transition 
(Deguchi \& Watson 1990). By integrating the radiative transfer equations
that are based 
on the density matrix formulation, Nedoluha \& Watson (1990a) finally obtained 
the full solution---the fractional linear polarization for all $R$---for
the GKK 
idealization. Complete agreement with what is expected 
from GKK was obtained. Answers to the question of how strong must the 
inequalities be in order for the GKK solutions to be applicable, and how 
does the linear polarization behave when they are not valid, were 
obtained. Representative results are reproduced in Figure 5. Note that the 
direction of the linear polarization is neither parallel nor perpendicular 
to the projection of the magnetic field on the plane of the sky when $R 
\simeq g\Omega$ (i.e., Stokes U$\not=$0). 
\begin{figure} 
\begin{center} 
\epsfig{file=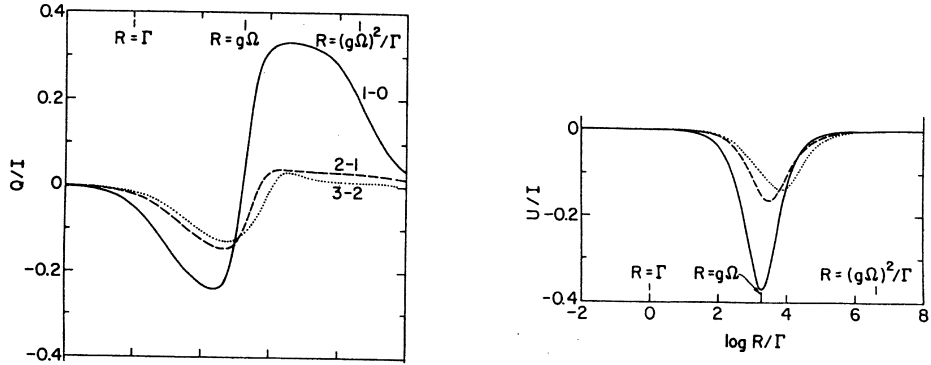, width=5in} 
\caption{\footnotesize Fractional linear polarizations Stokes Q/I and U/I
as a function of the degree of saturation for a linear maser in the regime
in which the Zeeman splitting is much smaller than the spectral
linebreadth. The results of calculations for angular momentum J=1-0, 2-1,
and 3-2 masing transitions are shown. The angle between the direction of
the maser beam and the magnetic field is 45 degrees. As indicated in the
figures, the ratio $g\Omega/\Gamma$= 2000 is adopted in this example. From
Nedoluha \&
Watson (1990a).} 
\end{center} 
\end{figure}

\vskip.2in
\noindent {\bf B.\ \ Additional Considerations for the Linear Polarization}
\vskip.1in
As noted  above, the GKK mechanism alone of saturated masing 
in the presence of a magnetic field is unlikely to produce the highest 
fractional linear polarizations that are observed in the J=2-1 and higher 
transitions of the circumstellar SiO masers.

Actual masing environments in astronomy may be more complicated in 
essential ways than the idealized environment considered by GKK. Of  these, 
anisotropic (or $m$-dependent) pumping of the magnetic substates is likely
to be 
the most important. The absorption and/or the escape of infrared radiation 
by the masing molecules ordinarily is essential in the pumping of 
astrophysical masers. If the angular distribution of this radiation is 
anisotropic when it is incident upon the molecules, or if the angular 
distribution with which it escapes is anisotropic, then the magnetic 
substates of a molecular energy level will be pumped unequally. Explicit 
calculations for the radiative transfer of anisotropic, infrared radiation 
that is involved in the pumping were performed in the context of the 
circumstellar SiO masers to demonstrate that sufficient anisotropy can be 
present under conditions where the magnitude of the pumping is adequate to 
provide the maser power that is observed (Western \& Watson 1983). The 
circumstellar SiO masers are natural candidates for anisotropic pumping 
since they are located within a few stellar radii of the star. In addition, 
the medium in which they are located probably has strong velocity gradients 
which can create preferential directions for the escape of the infrared 
radiation. As a group, the circumstellar SiO masers do exhibit the highest 
fractional linear polarizations.

Even though anisotropic pumping associated with infrared radiation may 
be the dominant cause for the linear polarization, the direction of the 
polarization vector is still likely to be either parallel or perpendicular 
to the direction of the magnetic field as projected onto the surface of the 
sky (Nedoluha \& Watson 1990a). This will occur as long as $g\Omega$ is 
somewhat greater than $R$ and $\Gamma$ (except possibly when $R$ and $\Gamma$ 
are comparable) because the magnetic field is the dominant axis for 
symmetry. An analogous effect occurs in the alignment of interstellar dust 
grains. They are aligned either parallel or perpendicular to the magnetic 
field, regardless of the mechanism that causes the alignment, because the 
dust grains precess rapidly around the direction of the magnetic field. 
Faraday rotation within the maser can, however, reduce the
magnitude and 
change the direction of the linear polarization just as it does for 
ordinary radiation (see Wallin \& Watson 1997 for an explicit calculation).
\vskip.2in 
\noindent {\bf C.\ \ Circular Polarization}
\vskip.1in
GKK considered only the polarization characteristics of the radiation at 
the center of the spectral line. In the weak splitting limit, no circular 
polarization occurs at line center.

The issues  can
be discussed in two ways. Firstly, to what degree is Stokes-V due to what 
we consider to be the standard Zeeman effect altered by the saturation of a 
maser. Secondly, what are ``non-Zeeman" effects that cause circular 
polarization. The distinction between these may not always be clear, in 
principle. In practice, it seems natural that ``alterations to the standard 
Zeeman effect" should refer to changes that can be described in terms of 
the unequal populations of the magnetic substates that are created in 
saturated masers in the presence of a magnetic field when $\Gamma\ll R\ll 
g\Omega$. Non-Zeeman effects encompass everything else.

\noindent {\it C1. Alteration of Zeeman circular polarization}

Initial calculations of the influence of saturation on the standard Zeeman 
effect focussed on the 22 GHz water transition (Nedoluha \& Watson 1992). 
The calculations were limited to saturations 
$R/\Gamma \leq 10$. They nevertheless demonstrated that the standard 
relationship ${\rm V}/\partial I/\partial\omega = {\rm constant} \times {\rm 
B cos\ }\theta$ 
 for two-level transitions, which is applicable for thermal 
spectral lines and for masing lines that are unsaturated, is altered by the 
effects of saturation. At larger angles $\theta,\ {\rm V}/(\partial 
I/\partial\omega)$, was found be greater by a factor of 2 or so for the 
F=7-6 hyperfine transition than would be expected from the standard 
relationship. Since the 22 GHz transition 
probably is a result of the merger of three hyperfine components, the 
standard relationship between V and $\partial I/(\partial\omega)$ most likely
is  
meaningless  Alternative relationships exist (e.g., Watson \& Wiebe 
2001). More extensive calculations have now been performed (Watson \& 
Wyld, in preparation) for J=1-0 and J=2-1 masing transitions. The quantity
V/$\partial I/(\partial\omega)$ is 
increased in these calculations beyond that given by the standard 
relationship by the effects of saturation. The magnitude of the increase is 
an increasing function of the angle $\theta$ until a maximum is reached at 
some large angle which depends somewhat on the degree of saturation. For 
$R/\Gamma\gtrsim 3, {\rm V}/(\partial I/\partial\omega)$ {\it increases, not 
decreases}, with angle $\theta$ within this range of angles.

\noindent {\it C2. Non-Zeeman circular polarization}

Modifications to the standard Zeeman effect, which have been 
discussed above, are likely to cause an uncertainty of only  a factor of a 
few in the inferred magnetic fields. In contrast, when they are effective, 
``non-Zeeman" mechanisms tend to create large circular polarization that is 
difficult to relate in a useful way to the strength of the magnetic field. 
Non-Zeeman circular polarization, as defined in the foregoing, occurs when 
the medium is anisotropic and the principal optical axes of the medium are 
not aligned with the direction of the linear polarization of radiation that 
is passing through the medium. The effect is analogous to the creation of 
circularly polarized radiation when ordinary, linearly polarized light 
passes through a sheet of polaroid material (or a ``quarter-wave plate") 
with axes that are not aligned with the direction of the polarization of 
the light. In astronomy, this effect is observed to occur for optical light 
when the direction of the magnetic field, and hence the direction of 
alignment of the polarizing dust grains, changes along the line of sight. 
The two requirements are then---firstly, the masing medium must be optically
anisotropic and secondly, its optical axes and the direction of the linear 
polarization of the radiation must be misaligned along the path of the 
radiation.

The masing medium will be anisotropic when the populations of the 
magnetic substates are unequal. They will be unequal when there is 
anisotropic pumping or when the linear 
polarization is created in the GKK mechanism. More generally, when linearly 
polarized radiation is created within a medium, it follows that the medium 
is anisotropic. The medium must be anisotropic to distinguish one direction 
from another---which is required to establish a direction for the linear 
polarization of the radiation. When linearly polarized radiation is 
observed, it then follows immediately that the first requirement for 
non-Zeeman circular polarization is satisfied unless the masing medium is 
simply amplifying background radiation that is already linearly polarized 
to the same degree that is observed. The masing medium is surely 
optically anisotropic in the spectral lines of the SiO masers 
which tend to be have high linear polarization.

Establishing when the second requirement for non-Zeeman 
circular polarization is satisfied is more difficult. When $g\Omega \gg R\ 
{\rm and}\ \Gamma$, the optical axes of the medium ordinarily will be 
aligned with the magnetic field. Thus, changes in the projected direction 
of the magnetic field on the plane of the sky within the masing gas will 
cause the optical axes and the direction of the linear polarization of the 
maser radiation to become misaligned. Using a statistical description of 
turbulent velocity and magnetic fields, we (Wiebe \& Watson 1998) have 
performed explicit computations to see what can be expected when
the path lengths for masing and the changes in the direction of the
magnetic field within 
these path lengths are determined by a plausible 
description of the medium. 
The characteristics of the linear and circular polarization that 
emerge in this calculation are similar to what is observed for the 
circumstellar SiO masers. 
Even in the absence of any change in the magnetic field, the 
optical axes and the direction of the linear 
polarization will become misaligned when $R\approx g\Omega$. This can 
easily be seen to occur, for example in Figure 5, where the direction of 
the linear polarization due to the GKK mechanism rotates by 90 degrees as $R$ 
approaches and significantly exceeds $g\Omega$. Creation of non-Zeeman 
circular polarization when $R\approx g\Omega$ has been investigated in 
considerable detail (Nedoluha \& Watson 1990b; 1994).

The conversion of linear to circular polarization in an anisotropic medium 
occurs because 
the phase velocity of a linearly polarized electromagnetic wave is 
different when its electric field is parallel to one, as opposed to the
other, of the 
optical axes. Within a spectral line, however, which of the axes has the 
larger phase velocity is reversed at the center of the spectral line. 
Radiative transitions of molecules are analogous to classical resonances in 
making a contribution to both the real and imaginary parts of the 
dielectric constant of the medium (e.g., Jackson 1999). The imaginary part 
of the dielectric constant describes the absorption and emission due to the 
radiative transition. The contribution of the resonance to the real part of 
the dielectric constant determines the phase velocity of the medium at 
frequencies within the spectral line. This contribution is antisymmetric 
about line center, and hence a reversal occurs at line center as to which
phase 
velocity is the larger. As a result, an antisymmetric Stokes-V is obtained 
when circularly polarized radiation is created by the non-Zeeman mechanisms 
being described here. The profile of this Stokes-V is quite similar to the 
profile that is expected when the Stokes-V is due to the standard Zeeman 
effect---at least for isolated, two-level transitions (e.g., Wiebe \& Watson 
1998 for representative spectra).

\vspace{12pt}
Finally, assertions have been made that "fluctuations", "instabilities" and
perhaps other ill defined, similar effects cause the polarization of masers
to be quite
different 
from the description in this Appendix. In all cases where such effects have
been 
examined in a meaningful way for astrophysical masers, 
they are unimportant (GKK; Western 1983;
Watson 1994;
Wallin \& Watson 1995). Non-local instabilities do exist in the radiative
transfer 
equations for astrophysical masers, but only in a quite limited region of
parameter space
(Scappaticci \& Watson 1992).
Further, in alternatives that have been proposed to the GKK/Illinois description 
of maser polarization, there is a fundamental failure to recognize that magnetic 
substates are no longer ``good'' or diagonal quantum states when $R$ 
approaches and exceeds $g\Omega$. Density matrix methods are utilized 
for quantal calculations in this regime, and are employed by in the GKK/Illinois 
calculations.

\end{document}